\documentclass[aip,pre,amsmath,amssymb,reprint,]{revtex4-1}

\usepackage{
    graphicx,
    bm
}

\begin{document}

\title{Conservative dissipation: How important is the Jacobi identity in the dynamics? }

\author{C.E. Caligan}
\email{ccaligan@gatech.edu}
\affiliation{
School of Physics, Georgia Institute of Technology, Atlanta, Georgia 30332-0430, USA
}

\author{C. Chandre}
 \email{chandre@cpt.univ-mrs.fr}
\affiliation{ 
Centre de Physique Th\'eorique, CNRS / Aix-Marseille Universit\'e, Campus de Luminy, 13009 Marseille, France
}

\date{\today}

\begin{abstract}
Hamiltonian dynamics are characterized by a function, called the Hamiltonian, and a Poisson bracket. The Hamiltonian is a conserved quantity due to the anti-symmetry of the Poisson bracket. The Poisson bracket satisfies the Jacobi identity which is usually more intricate and more complex to comprehend than the conservation of the Hamiltonian. Here we investigate the importance of the Jacobi identity in the dynamics by considering three different types of conservative flows in ${\mathbb R}^3$: Hamiltonian, almost-Poisson and metriplectic. The comparison of their dynamics reveals the importance of the Jacobi identity in structuring the resulting phase space.   
\end{abstract}


\maketitle

\textbf{
Hamiltonian systems are ubiquitous in many branches of physics, e.g., in atomic and molecular physics, in celestial mechanics, and in fluid and plasma physics, to name a few. These systems display a finite or infinite number of degrees of freedom depending on the complexity and modeling of the problem at hand. Hamiltonian systems have the important property of being conservative, in the sense that there exists a conserved quantity, which is most often the energy, called the Hamiltonian, and denoted by $H$ in what follows. However Hamiltonian systems are much more than conservative systems. Hamiltonian dynamics is generated by a Poisson bracket $\{\cdot,\cdot\}$. The time-evolution of a given observable $F$ (function of the dynamical variables) is given by 
$$
\frac{dF}{dt}=\{F,H\}.
$$
This Poisson bracket is antisymmetric, bilinear and satisfies the Leibniz rule and the Jacobi identity. The antisymmetry property, i.e., $\{F_1,F_2\}=-\{F_2,F_1\}$, implies that the Hamiltonian is a conserved quantity, i.e., $dH/dt=0$. The richness of Hamiltonian systems resides in the Jacobi identity~:
$$
\{F_1,\{F_2,F_3\}\}+\{F_2,\{F_3,F_1\}\}+\{F_3,\{F_1,F_2\}\}=0,
$$
for all observables $F_1$, $F_2$ and $F_3$. In general, this identity makes it extremely difficult to deal with Hamiltonian systems, especially for non-canonical Hamiltonian systems~\cite{morr98}. For instance, performing approximations on the equations of motion, e.g., neglecting terms, usually breaks up this property if not done carefully.  
The natural question is how important is this property on the actual dynamics. Can we disregard this identity without affecting qualitatively the dynamics? Is it sufficient for a good model to be energy conserving in absence of dissipative terms? }

Numerous conservative models in fluid and plasma physics are built without considering the Jacobi identity. In the field of nonlinear control theory, some works in the literature got rid of this property, and coined the resulting systems, almost-Poisson (or pseudo-Poisson). These almost-Poisson systems are conservative systems (and the corresponding bracket satisfies the Leibnitz rule). However, due to the fact that the Jacobi identity is not satisfied, there is some dissipation associated with the almost-Poisson systems, which has been coined ``fake dissipation'' in Ref.~\cite{morr86}. This fake dissipation enters in competition with the dissipative terms introduced in the equations of motion. For instance, models in fluid and plasma dynamics typically exhibit some dissipation modeling the interaction of the considered degrees of freedom with the ones which have been left out to derive a reduced (and hence more tractable) model. For instance, the effect of neglected degrees of freedom, e.g., associated with small scales, can be modeled by diffusion terms in the equations of motion. In all these models, the ideal part which is obtained by removing the dissipative terms (e.g., the ones characterized by phenomenological constants, like diffusion coefficients, viscosity, collisionality...) should be Hamiltonian, reflecting the Hamiltonian character of the parent models (or from first principles) from which the reduced models have been constructed (e.g., Vlasov-Maxwell equations in plasma physics). Is it good enough to have an almost-Poisson model with some added dissipation? Or even in the dissipative case, is it still important to have the Jacobi identity for the ideal part of the model?  

The present article provides numerical evidence using a simple example for the importance of the Jacobi identity in the dynamics, which drastically affect the qualitative dynamics. 

Here we consider flows in ${\mathbb R}^3$ given by
the bracket
$$
\{F,G\}={\bm \alpha}\cdot \nabla F\times \nabla G,
$$
where ${\bm \alpha}={\bm \alpha}(x,y,z)$ is a function from ${\mathbb R}^3$ to ${\mathbb R}^3$.
This bracket is bilinear and antisymmetric. Moreover it satisfies the Leibniz rule for any function ${\bm\alpha}$. It satisfies the Jacobi identity if and only if
\begin{equation}
\label{eq:jac3}
{\bm\alpha}\cdot\nabla\times {\bm\alpha}=0.
\end{equation}
We notice that Nambu systems~\cite{namb73} correspond to the case where ${\bm \alpha}=\nabla S $. In this case, the velocity $\dot{\bm x}=\nabla H\times {\bm \alpha}$ is divergence-free. This is not the case for all ${\bm \alpha}$ satisfying Eq.~(\ref{eq:jac3}).  
We recall that Casimir invariants are defined as observables which Poisson-commute with all the other observables. In particular it commutes with the Hamiltonian, and is therefore a conserved quantity. In our case, a Casimir invariant $S(x,y,z)$ has to satisfy $\alpha\times \nabla S=0$. Therefore there exists a function $\lambda(x,y,z)$ such that ${\bm \alpha}=\lambda \nabla S$ from which it can be deduced that ${\bm \alpha}\cdot\nabla \times {\bm \alpha}=0$. As a consequence, if the system possesses a Casimir invariant, the Jacobi identity is automatically satisfied. Moreover, if has been proved in Ref.~\cite{ay03} that there exists two scalar functions $\lambda$ and $S$ such that the solution of Eq.~(\ref{eq:jac3}) can be written locally as ${\bm\alpha}=\lambda \nabla S$. Thus $S$ is a Casimir invariant. 
The Jacobi identity is locally equivalent to the existence of a Casimir invariant for flows in ${\mathbb R}^3$. If the system is almost-Poisson, there is no additional conserved quantity in general. 

In order to illustrate its impact on the dynamics we consider Hamiltonian \( H\) and ${\bm \alpha}$ given by
\begin{eqnarray}
    && H(x,y,z) = \cos(x+y+z),\label{eq:ham} \\
    && {\bm\alpha} = (\cos x,\cos y, \sin z+\varepsilon \cos y).
\end{eqnarray}
We notice that for $\varepsilon=0$ the system is Hamiltonian with a Casimir invariant 
$$
S(x,y,z)=\sin x+\sin y -\cos z,
$$
whereas for $\varepsilon \not= 0$, it is almost-Poisson since ${\bm\alpha}\cdot\nabla\times {\bm\alpha}=-\varepsilon \cos x \sin y$. 

\begin{figure}
\centering
\includegraphics[scale=0.9]{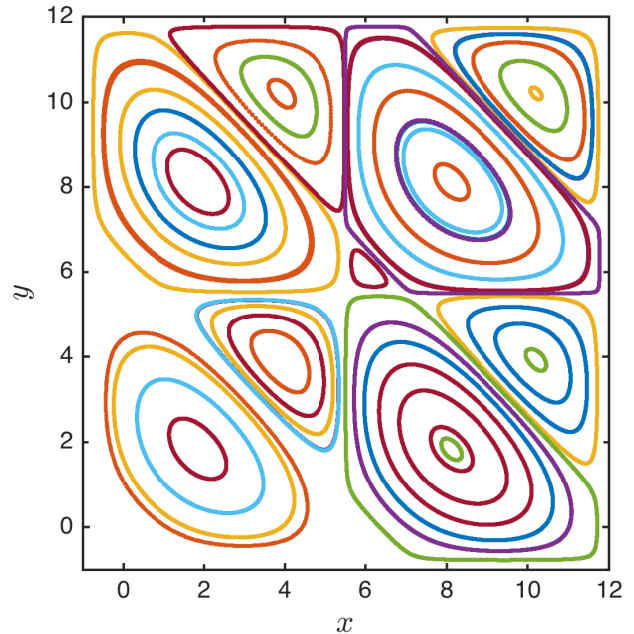}
\caption{Trajectories associated with Eqs.~(\ref{eq:C1_1}-\ref{eq:C1_3}) with $\epsilon=0$. This corresponds to a Hamiltonian case.}
\label{fig:Ham}
\end{figure}

\begin{figure}
\centering
\includegraphics[scale=0.9]{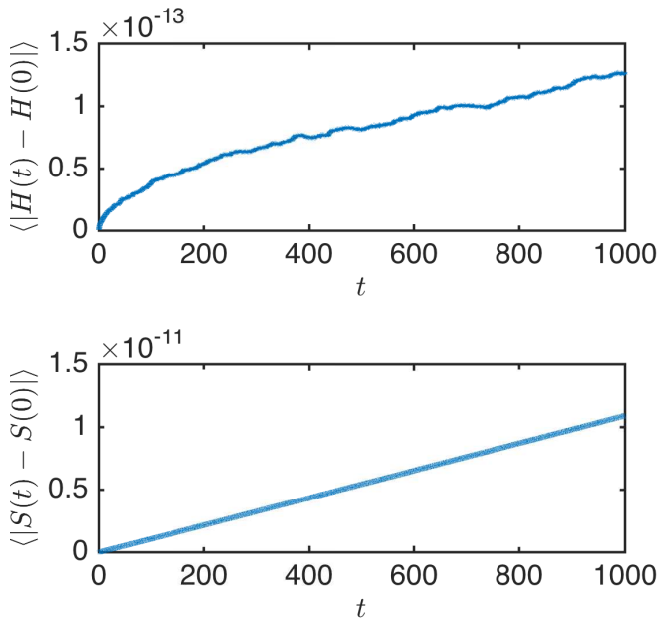}
\caption{Mean value of the energy variations (upper panel) and Casimir variations (lower panel) for  System (\ref{eq:C1_1}-\ref{eq:C1_3}) with $\epsilon=0$. This corresponds to a Hamiltonian case.}
\label{fig:Ham_CQ}
\end{figure}

We consider a large ensemble of initial conditions on the plane $x+y+z=\pi/4$ and we integrate these trajectories for a rather long time (up to $T=1000$) to observe the effect of dissipation. We monitor the conservation of the energy $H$ and the Casimir invariant $S$.  The equations of motion are 
\begin{subequations}
\begin{align}
& \dot{x}=-\sin(x+y+z)(\sin z-\cos y)\nonumber \\
& \qquad \qquad \qquad \qquad -\varepsilon \cos y \sin(x+y+z), \label{eq:C1_1}\\
& \dot{y}=-\sin(x+y+z)(-\sin z+\cos x) \nonumber \\
& \qquad \qquad \qquad \qquad +\varepsilon \cos y \sin(x+y+z),\label{eq:C1_2}\\
& \dot{z}=-\sin(x+y+z)(\cos y-\cos x). \label{eq:C1_3}
\end{align}
\end{subequations}
The phase portrait is depicted in Fig.~\ref{fig:Ham}. As expected, we notice that energy and the Casimir invariant, depicted in Fig.~\ref{fig:Ham_CQ}, are conserved up to numerical precision. Almost all trajectories are periodic given the fact that there are two conserved quantities. It is worth noting that the shape of the phase portrait, and in particular the absence of attractors, is often attributed to the fact that the flow is divergence-free (or volume-preserving), which is the case here. However by choosing another function ${\bm \alpha}=\lambda \nabla S$, the phase portrait is exactly the same (even if the trajectories on these one-dimensional curves are different) since these one-dimensional curves are defined by the two same conserved quantities $H$ and $S$. 

\begin{figure}
\centering
\includegraphics[scale=0.9]{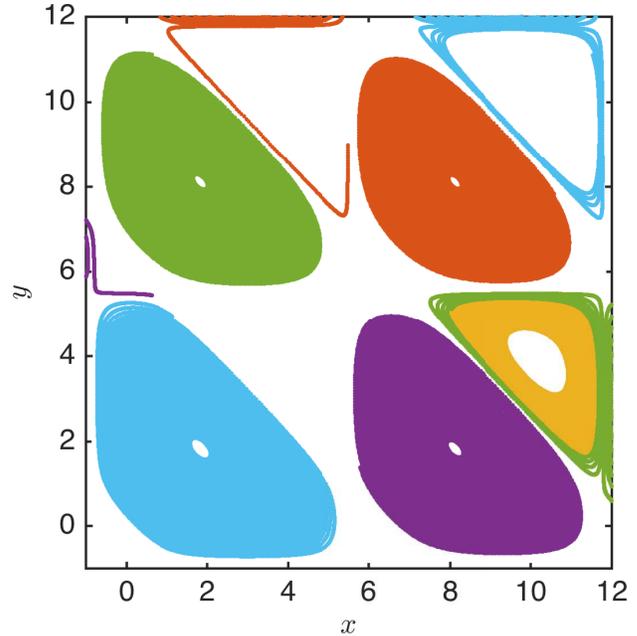}
\caption{Trajectories of System~(\ref{eq:C1_1}-\ref{eq:C1_3}) with $\epsilon=10^{-2}$. This corresponds to an almost-Poisson case.}
\label{fig:aP}
\end{figure}

\begin{figure}
\centering
\includegraphics[scale=0.9]{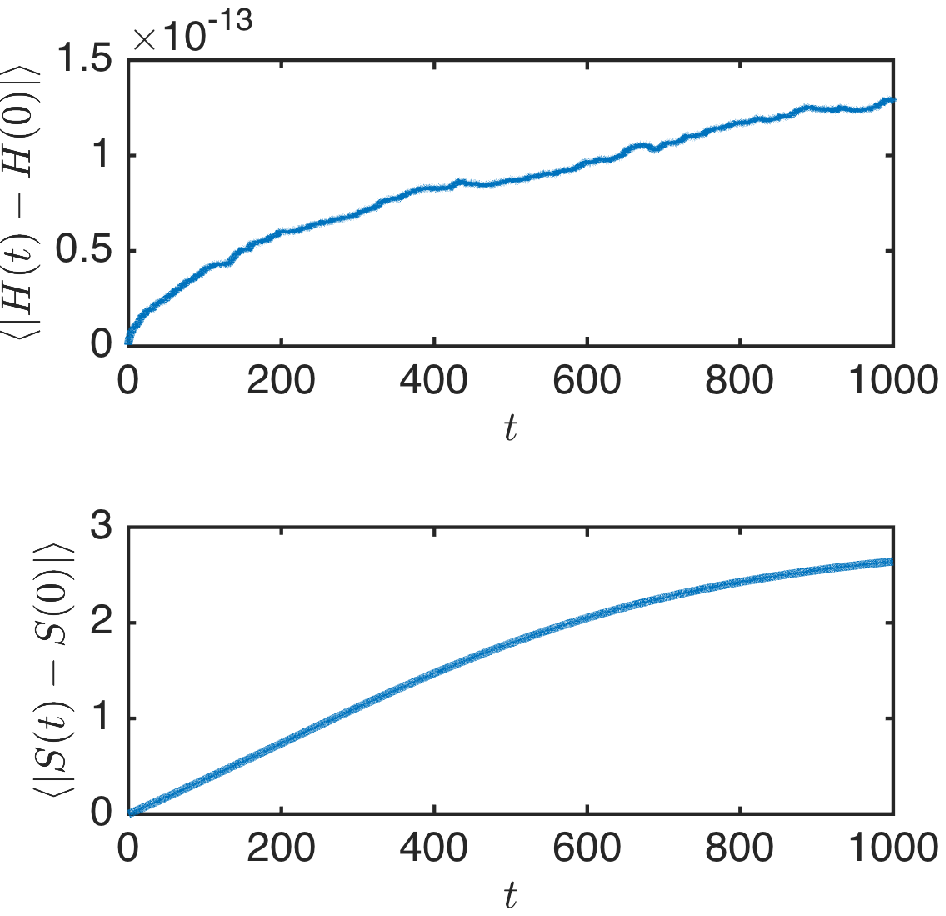}
\caption{Mean value of the variations of $H$ (upper panel) and $S$ (lower panel) for  System~(\ref{eq:C1_1}-\ref{eq:C1_3}) with $\epsilon=10^{-2}$. This corresponds to an almost-Poisson case.}
\label{fig:aP_CQ}
\end{figure}

\begin{figure}
\centering
\includegraphics[scale=0.9]{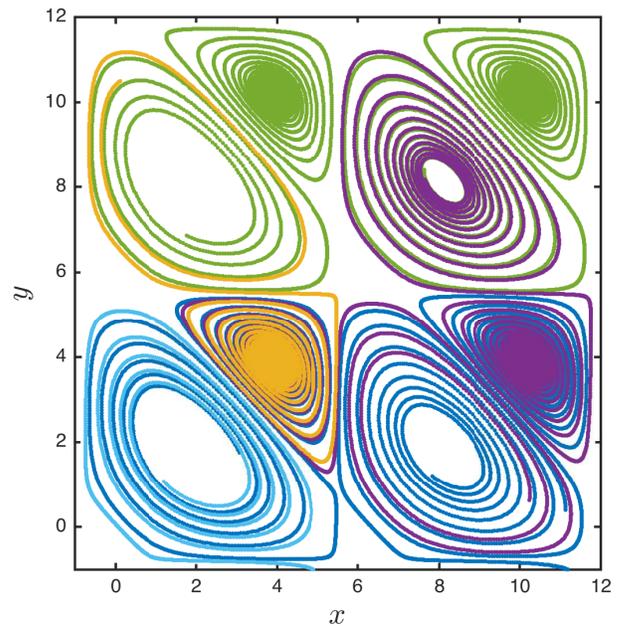}
\caption{Trajectories of System~(\ref{eq:C2_1}-\ref{eq:C2_3}) with $\bar{\mu}=10^{-2}$. This corresponds to a metriplectic case.}
\label{fig:mp}
\end{figure}

\begin{figure}
\centering
\includegraphics[scale=0.9]{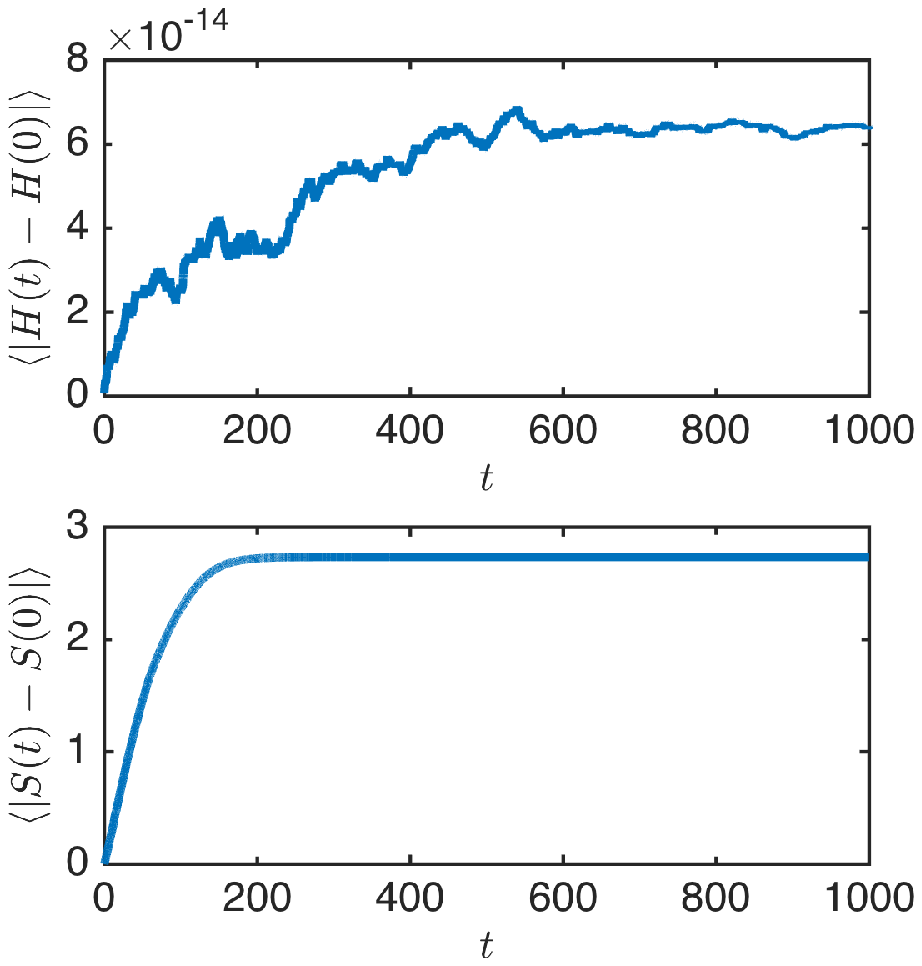}
\caption{Mean value of the variations of $H$ (upper panel) and $S$ (lower panel) for System (\ref{eq:C2_1}-\ref{eq:C2_3}) with $\bar{\mu}=10^{-2}$. This corresponds to a metriplectic case.}
\label{fig:mp_CQ}
\end{figure}

We compare the Hamiltonian dynamics $\epsilon=0$ with another conservative system obtained from Eqs.~(\ref{eq:C1_1}-\ref{eq:C1_3}) with a slight modification, $\epsilon=10^{-2}$. Typical trajectories are plotted in Fig.~\ref{fig:aP}, and the associated variation of energy and Casimir invariant are plotted in Fig.~\ref{fig:aP_CQ}.
First we notice that the energy is conserved up to numerical precision. In addition, we clearly notice the presence of attractors even though this system is conservative. As expected, $S$ is no longer a conserved quantity. For individual trajectories, $S(t)$ is not monotonous, even if, on average, it appears to be monotonous. However here the convergence towards attractors is very slow, which explains why the trajectories appear to fill densely some parts of phase space. 
We compared the dissipative dynamics ($\varepsilon \not= 0$) to another type of dissipation, a metriplectic system \cite{morr86,Tfish05,bloc13}. The idea is to construct a conservative system with an entropy. We define this system from the following bracket
$$
(F,G)={\bm \alpha}\cdot \nabla F\times \nabla G+\mu (\nabla F\times \nabla H)\cdot (\nabla H\times \nabla G),
$$
where \(\mu=\mu(x,y,z)\) is a positive function which will be chosen later.
The bracket $(\cdot, \cdot)$ is bilinear but obviously not a Poisson bracket since it is not antisymmetric. In fact the part that has been added to the antisymmetric bracket $\{\cdot,\cdot\}$ is symmetric. We assume that $\{\cdot,\cdot\}$ is a Poisson bracket, and hence it has a Casimir invariant, denoted by $S(x,y,z)$. We define the free energy as $F_0=H-S$.  We define the dynamics as
$$
\frac{dF}{dt}=(F,F_0). 
$$
We notice that $H$ is a conserved quantity by construction ($S$ being a Casimir invariant of the Poisson part of the bracket). Moreover, the dynamics of $S$ verifies
$$
\frac{dS}{dt}=\mu \Vert \nabla S \times \nabla H\Vert^2\geq 0,
$$
which means that the entropy grows with time monotonically. 
In the numerical example, we consider $\mu=\bar{\mu}/\sin^2(x+y+z)$ where $\bar{\mu}$ is a constant. The equations of motion for the metriplectic systems become:
\begin{subequations}
\begin{align}
& \dot{x}=-\sin(x+y+z)(\sin z-\cos y)\nonumber \\
& \qquad \qquad \qquad +\bar{\mu}(-2\cos x+\cos y +\sin z), \label{eq:C2_1}\\
& \dot{y}=-\sin(x+y+z)(-\sin z+\cos x) \nonumber\\
& \qquad\qquad\qquad+\bar{\mu}(\cos x-2\cos y+\sin z),\label{eq:C2_2}\\
& \dot{z}=-\sin(x+y+z)(\cos y-\cos x)\nonumber \\
& \qquad\qquad\qquad+\bar{\mu}(\cos x+\cos y -2\sin z). \label{eq:C2_3} 
\end{align}
\end{subequations}
The system is conservative in the sense that $H=\cos(x+y+z)$ is a conserved quantity. It is dissipative in the sense that it is not Hamiltonian for $\bar{\mu}\not= 0$.  The phase portrait is depicted in Fig.~\ref{fig:mp} and the values of the variations of $H$ and $S$ are represented on Fig.~\ref{fig:mp_CQ}. We emphasize that the integration time for System~(\ref{eq:C2_1}-\ref{eq:C2_3}) is the same as the one for System~(\ref{eq:C1_1}-\ref{eq:C1_3}). However both types of dissipation result in significantly different phase portraits. The dissipation introduced in the metriplectic system is much stronger leading to a fast convergence towards attracting fixed points. This is also seen in Fig.~\ref{fig:mp_CQ} where the convergence of the mean value of the variations of $S$ is much stronger in the metriplectic system than in the almost-Poisson case. 


The comparison of the numerics shows that, even though the energy is conserved to a high accuracy, the dynamics exhibited by the two dissipative systems, built from the same ideal part, are very different. This reinforces the importance of the type dissipative terms introduced in the equations.  

In summary, we have shown on a simple example, Hamiltonian flows in ${\mathbb R}^3$, that the Jacobi identity shapes the dynamics by preventing the existence of attractors. For instance, we have emphasized the strong links between the Jacobi identity and the existence of Casimir invariants. These invariants foliate phase space, preventing unphysical transport across phase space. Despite the fact that systems are energy-conserving, Hamiltonian systems exhibit a qualitatively different dynamics that almost-Poisson systems or metriplectic systems. Of course, these results are particular to Hamiltonian flows in ${\mathbb R}^3$, but the link between Casimir invariants has also been noticed for infinite dimensional Hamiltonian systems, for instance, for fluid reductions of the Vlasov equation, and for Dirac brackets of constrained Hamiltonian systems \cite{peri15,chan13}.

\begin{acknowledgments}

The research leading to these results has received funding from the People Program (Marie Curie Actions) of the European Union's Seventh Framework Program No. FP7/2007-2013/ under REA Grant No. 294974. CC acknowledges useful discussions with P.J. Morrison.

\end{acknowledgments}

\end{document}